\documentclass[preprintnumbers,amsmath,amssymbm,prd]{revtex4}
\usepackage{epsfig}
\usepackage{graphicx}
\usepackage{amssymb}

\begin{document}
\title{Electroweak clouds supported by magnetically charged black holes: Analytic treatment 
along the existence-line}
\author{Shahar Hod}
\affiliation{The Ruppin Academic Center, Emeq Hefer 40250, Israel}
\affiliation{ }
\affiliation{The Hadassah Institute, Jerusalem 91010, Israel}
\date{\today}

\begin{abstract}
\ \ \ It has recently been revealed [R. Gervalle and M. S. Volkov, Phys. Rev. Lett. (2024)] 
that magnetically charged black holes of the composed Einstein-Weinberg-Salam field theory 
can support bound-state hairy configurations of electroweak fields. 
In the present paper we study, using {\it analytical} techniques, 
the physical and mathematical properties of the supported 
linearized electroweak fields (spatially regular electroweak 'clouds') 
in the dimensionless large-mass $\mu\equiv m_{\text{w}} r_+\gg1$ regime  of the 
composed black-hole-field system (here $m_{\text{w}}$ is 
the mass of the supported W-boson field and $r_+$ is the outer horizon radius of the central 
supporting black hole). 
In particular, we derive a remarkably compact formula for
the discrete resonance spectrum
$\{\mu_k(n)\}_{k=0}^{k=\infty}$ that characterizes the composed black-hole-linearized-field configurations, 
where the integer $n\equiv 2Pe\in\mathbb{Z}$ characterizes the discrete charge parameter $P$ of the central 
magnetic Reissner-Nordstr\"om black hole and $e$ is the electron charge. 
The physical significance of the analytically derived resonant spectrum
stems from the fact that, in the dimensionless large-charge $n\gg1$ regime, 
the fundamental (largest) eigenvalue $\mu_0(n)$ determines the critical existence-line of
the composed Einstein-Weinberg-Salam theory, a boundary line that 
separates hairy magnetic-black-hole-electroweak-field bound-state 
configurations from bald magnetic Reissner-Nordstr\"om black holes. 
\end{abstract}
\bigskip
\maketitle

\section{Introduction}

Matter and radiation fields in asymptotically flat black-hole spacetimes are
expected, according to the no-hair conjecture \cite{Whee,Car}, to be absorbed by the central 
black hole or to be scattered away to infinity. 
Early explorations of the Einstein-matter field equations \cite{Chas,Hart,BekVec} (see also \cite{BekMay,Her1n,Hodstationary}) have revealed 
the fact that, in accord with this influential conjecture, black holes cannot support 
spatially regular static matter configurations which are made of scalar fields, 
spinor fields, and massive vector fields.

The first physically interesting counter-example to the no-hair conjecture was provided in \cite{fir}, 
where it was explicitly proved that asymptotically flat hairy black-hole solutions exist in the 
composed Einstein-Yang-Mills field theory (see \cite{VolGal} for a 
review and a detailed list of references). 
Subsequent studies have revealed the existence of black-hole spacetimes with a Skyrme
hair \cite{hairSky}, a Higgs hair \cite{hairHig1,hairHig2}, stringy hair \cite{hairstr}, 
a stationary scalar hair \cite{Hod1,Hod2,Her1,Her2,Gar1}, and a scalar hair non-minimally 
coupled to the Maxwell electromagnetic invariant \cite{Hersc1,Hersc2,Hodsc1,aaa1,aaa2,Moh}.

Intriguingly, it has recently been revealed in the physically 
important work \cite{GV} that, in the composed Einstein-Weinberg-Salam field theory, 
spatially regular electroweak field configurations can be supported in curved spacetimes of 
magnetically charged black holes. 
In particular, the numerical results presented in \cite{GV} have revealed that the 
Einstein-Weinberg-Salam theory is characterized by the existence of a charge-dependent 
critical {\it existence-line} $m_{\text{w}} r_+=m_{\text{w}} r_+(n)$ that marks
the boundary between composed magnetic-black-hole-electroweak-field hairy 
configurations and bald magnetic Reissner-Nordstr\"om black holes. 
Here $r_+$ is the outer horizon radius of the central supporting black hole, 
$P=n/(2e)$ is the discrete charge parameter of the magnetic black hole 
[where $n\in\mathbb{Z}$ and $e$ is the electron charge, see Eq. (\ref{Eq4}) below], 
and $m_{\text{w}}$ is the mass of the W-boson. 

The critical existence-line of the composed Einstein-Weinberg-Salam field theory corresponds to spatially
regular linearized electroweak field configurations that are supported by a
central magnetically charged black hole with a Reissner-Nordstr\"om geometry (the term `linearized cloud' 
is usually used in the physics literature \cite{Hod1,Hod2,Her1,Her2,Gar1} 
in order to describe a supported bound-state field configuration that sits on the critical existence-line 
of the composed black-hole-field system).

The main goal of the present compact paper is to study, using {\it
analytical} techniques, the physical and mathematical properties of
the composed magnetic-Reissner-Nordstr\"om-black-hole-linearized-electroweak-field 
bound-state configurations in the dimensionless large-mass regime $m_{\text{w}} r_+\gg1$. 
In particular, below we shall derive a remarkably compact analytical
formula [see Eq. (\ref{Eq38}) below] for the critical existence-line $m_{\text{w}} r_+=m_{\text{w}} r_+(n)$ that 
characterizes the composed Einstein-Weinberg-Salam field theory. 

\section{Description of the system}

We shall study the physical and mathematical properties of spatially regular 
linearized electroweak field configurations (massive electroweak clouds) which are supported 
by magnetically charged black holes. 
The composed Einstein-Weinberg-Salam field theory is characterized by a metric $g_{\mu\nu}$, 
an $SU(2)$ field $W=T_aW^a_{\mu}dx^{\mu}$ with $T_a=\tau_a/2$ where $\tau_a$ are the 
Pauli matrices, a $U(1)$ hypercharge field $B=B_{\mu}dx^{\mu}$, and a 
complex doublet Higgs field $\Phi$ \cite{GV}. 

The action of the composed physical system is given by the integral expression \cite{GV}
\begin{equation}\label{Eq1}
S={{e^2}\over{4\pi\alpha}}\int{\Big[{{1}\over{2\kappa}}R-{{1}\over{4g^2}}W^{a}_{\mu\nu}W^{a\mu\nu}- 
{{1}\over{4{g'}^2}}B_{\mu\nu}B^{\mu\nu}-(D_{\mu}\Phi)^+(D^{\mu}\Phi)-
{{\beta}\over{8}}(\Phi^+\Phi-1)^2\Big]\sqrt{-g}}d^4x\  ,
\end{equation}
where $R$ is the Ricci scalar, $W^{a}_{\mu\nu}$ is the $SU(2)$ gauge field strength, and 
$B_{\mu\nu}$ is the $U(1)$ gauge field strength. 
The values of the coupling constants are $\beta\simeq1.88$, $g=\cos\theta_{\text{w}}\simeq\sqrt{0.78}$, 
and $g'=\sin\theta_{\text{w}}\simeq\sqrt{0.22}$ \cite{GV}. 

The spacetime of the supporting magnetic Reissner-Nordstr\"om black hole 
is characterized by the curved line element \cite{GV}
\begin{equation}\label{Eq2}
ds^2=-f(r)dt^2+{1\over{f(r)}}dr^2+r^2(d\theta^2+\sin^2\theta
d\phi^2)\  .
\end{equation}
The dimensionless metric function in (\ref{Eq2}) is given by the 
radially-dependent expression \cite{GV,Noteunits}
\begin{equation}\label{Eq3}
f(r)=1-{{2M}\over{r}}+{{Q^2}\over{r^2}}\  ,
\end{equation}
where $M$ is the mass of the black hole. 
The magnetic charge of the black hole is given by the discrete relation \cite{GV,Notenp}
\begin{equation}\label{Eq4}
P=\sqrt{{{2}\over{\kappa}}}Q={{n}\over{2e}}\ \ \ \ ; \ \ \ \ n\in\mathbb{Z}\  ,
\end{equation}
where 
\begin{equation}\label{Eq5}
\kappa={{4e^2}\over{\alpha}}\Big({{m_{\text{z}}}\over{M_{\text{Pl}}}}\Big)^2\simeq 5.42\times10^{-33}\
\end{equation}
is the gravity coupling of the theory \cite{GV}. 
The electron charge and the fine structure constant are given respectively by the dimensionless 
expressions $e=gg'\simeq0.414$ and $\alpha\simeq1/137$ \cite{GV}. 
The boson masses are given by the relations \cite{GV} 
\begin{equation}\label{Eq6}
m_{\text{z}}=1/\sqrt{2}\ \ \ \ ;\ \ \ \ m_{\text{w}}=g m_{\text{z}}\ \ \ \ ;\ \ \ \ 
m_{\text{H}}=\sqrt{\beta}m_{\text{z}}
\end{equation}
in units of the energy (mass) scale $128.9$GeV. 

The horizon radii
\begin{equation}\label{Eq7}
r_{\pm}=M\pm(M^2-Q^2)^{1/2}\
\end{equation}
of the central supporting magnetic Reissner-Nordstr\"om black hole are
determined by the roots of the dimensionless metric function (\ref{Eq3}). 

As shown in \cite{GV}, the linearized field $w_{\mu}=(\delta W^1_{\mu}+i\delta W^2_{\mu})/g$ 
can be decomposed in the form 
\begin{equation}\label{Eq8}
w_{\mu}dx^{\mu}=e^{i\omega t}\psi(r)(\sin\theta)^{j+1}\sum_{m=-j}^{m=j}c_m z^{m-1}dz\  ,
\end{equation}
where $z=\tan(\theta/2)\exp(-i\varphi)$, $j=n/2-1$, and $n\geq2$. 
The radially-dependent function $\psi(r)$ satisfies the Schr\"odinger-like ordinary differential 
equation \cite{GV}
\begin{equation}\label{Eq9}
{{d^2\psi}\over{dy^2}}+(\omega^2-V)\psi=0\  ,
\end{equation}
where 
\begin{equation}\label{Eq10}
V=V[r(y)]=\Big(1-{{2M}\over{r}}+{{Q^2}\over{r^2}}\Big)\cdot\Big(m^2_{\text{w}}-{{n}\over{2r^2}}\Big)\
\end{equation}
is the effective radial potential of the composed magnetic-black-hole-electroweak-field bound-state 
configurations. 
The tortoise coordinate $y$ in (\ref{Eq9}) is defined by the dimensionless 
differential relation \cite{Notemap}
\begin{equation}\label{Eq11}
{{dr}\over{dy}}=f(r)\  .
\end{equation}
We shall henceforth consider static black-hole-linearized-electroweak-field bound-state configurations which are 
characterized by the relation $\omega=0$. 

The Schr\"odinger-like differential equation (\ref{Eq9}), supplemented 
the physically motivated boundary conditions
\begin{equation}\label{Eq12}
\psi(r=r_+)<\infty\
\end{equation}
and 
\begin{equation}\label{Eq13}
\psi(r\to\infty)\sim{{1}\over{r}}e^{-m_{\text{w}}r}\
\end{equation}
at the black-hole horizon and at spatial infinity, 
determine the resonant spectrum $\{\mu_k(n)\}_{k=0}^{k=\infty}$ of 
the composed magnetic-black-hole-linearized-electroweak-field bound-state configurations, 
where 
\begin{equation}\label{Eq14}
\mu\equiv m_{\text{w}} r_+\
\end{equation}
is the dimensionless mass parameter of the black-hole-field system. 

In the next section we shall explicitly prove that the discrete resonant 
spectrum $\{\mu_k(n)\}_{k=0}^{k=\infty}$, which characterizes the 
composed black-hole-field cloudy configurations of the 
Einstein-Weinberg-Salam theory (\ref{Eq1}), can be determined {\it analytically} 
in the dimensionless large-mass regime $\mu\gg1$. 

\section{The discrete resonant spectrum of the composed 
magnetic-black-hole-electroweak-field cloudy configurations: A WKB analysis}

In the present section we shall derive, using analytical techniques which are valid 
in the dimensionless large-mass regime
\begin{equation}\label{Eq15}
\mu\gg1\  ,
\end{equation}
a remarkably compact formula for the discrete resonant spectrum that characterizes the 
composed magnetically-charged-black-hole-linearized-electroweak-field bound-state configurations.

As we shall now show explicitly, the Schr\"odinger-like ordinary differential equation
(\ref{Eq9}), which determines the radial functional behavior of the 
bound-state electroweak field configurations in the magnetic black-hole spacetime (\ref{Eq2}), is
amenable to a WKB analysis in the dimensionless regime (\ref{Eq15}). In
particular, a standard second-order WKB analysis of the
Schr\"odinger-like equation (\ref{Eq9}) yields the mathematically compact 
discrete quantization condition \cite{WKB1,WKB2,WKB3}
\begin{equation}\label{Eq16}
\int_{y_{t-}}^{y_{t+}}dy\sqrt{-V(y;n)}=\big(k+{1\over2}\big)\cdot\pi\
\ \ \ ; \ \ \ \ k=0,1,2,...\  ,
\end{equation}
where the integration boundaries $\{y_{t-},y_{t+}\}$ in (\ref{Eq16}), 
which are determined by the functional relations 
\begin{equation}\label{Eq17}
V(y_{t-})=0\ \ \ \ \text{and}\ \ \ \ V(y_{t+})=0\  ,
\end{equation}
are the classical turning points of the composed magnetic-black-hole-electroweak-field radial potential (\ref{Eq10}). 
Here $k\in\mathbb{N}$ is the resonance parameter that 
characterizes the discrete resonant spectrum
$\{\mu_k(n)\}_{k=0}^{k=\infty}$ of the composed black-hole-field cloudy configurations.

Taking cognizance of the differential relation (\ref{Eq11}), one can
express the WKB resonance condition (\ref{Eq16}) in the integral form
\begin{equation}\label{Eq18}
\int_{r_{t-}}^{r_{t+}}dr{{\sqrt{-V(r;n)}}\over{f(r)}}=\big(k+{1\over2}\big)\cdot\pi\
\ \ \ ; \ \ \ \ k=0,1,2,...\  ,
\end{equation}
where the two turning points (radial integration boundaries) 
\begin{equation}\label{Eq19}
r_{t-}=r_+=M+(M^2-Q^2)^{1/2}
\end{equation}
and 
\begin{equation}\label{Eq20}
\gamma\equiv r_{t+}={{\sqrt{n}}\over{\sqrt{2}m_{\text{w}}}}\
\end{equation}
in (\ref{Eq18}) are determined by the radial relations [see Eqs. (\ref{Eq10}) and (\ref{Eq17})]
\begin{equation}\label{Eq21}
1-{{2M}\over{r_{t-}}}+{{Q^2}\over{r^2_{t-}}}=0\ \ \ \ \text{and}\ \ \ \ m^2_{\text{w}}-{{n}\over{2r^2_{t+}}}=0\  .
\end{equation}

Substituting Eqs. (\ref{Eq3}), (\ref{Eq10}), (\ref{Eq19}), and (\ref{Eq20}) into Eq. (\ref{Eq18}), 
one obtains the integral relation 
\begin{equation}\label{Eq22}
m_{\text{w}}\cdot\int_{r_+}^{\gamma}dr\sqrt{{{{{\gamma^2}\over{r^2}}-1}\over{1-{{2M}\over{r}}+{{Q^2}\over{r^2}}}}}
=\big(k+{1\over2}\big)\cdot\pi\
\ \ \ ; \ \ \ \ k=0,1,2,...\  .
\end{equation}

We shall now prove that the WKB resonance condition 
(\ref{Eq22}) is amenable to an {\it analytical} treatment in the dimensionless regime (\ref{Eq15}) of large 
black-hole-field masses. 
To this end, it is convenient to define the dimensionless physical quantities
\begin{equation}\label{Eq23}
x\equiv {{r-r_{\text{+}}}\over{r_{\text{+}}}}\ll1\  
\end{equation}
and
\begin{equation}\label{Eq24}
\epsilon\equiv {{\gamma}\over{r_+}}-1\ll1\  ,
\end{equation}
in terms of which one finds the dimensionless relation 
\begin{equation}\label{Eq25}
{{{{{\gamma^2}\over{r^2}}-1}\over{1-{{2M}\over{r}}+{{Q^2}\over{r^2}}}}}=
{{2r_+}\over{r_+-r_-}}\cdot\Big[{{\epsilon}\over{x}}-1+O(x,\epsilon)\Big]\  .
\end{equation}
Substituting Eqs. (\ref{Eq23}) and (\ref{Eq25}) into Eq. (\ref{Eq22}) one obtains the 
remarkably compact resonance condition
\begin{equation}\label{Eq26}
\mu\cdot
\int_{0}^{\epsilon}dx\sqrt{{{2r_+}\over{r_+-r_-}}\cdot\Big({{\epsilon}\over{x}}-1\Big)}
=\big(k+{1\over2}\big)\cdot\pi\
\ \ \ ; \ \ \ \ k=0,1,2,...\  .
\end{equation}

Defining 
\begin{equation}\label{Eq27}
z\equiv {{x}\over{\epsilon}}\
\end{equation}
one can express the WKB resonance condition (\ref{Eq26}) in the form 
\begin{equation}\label{Eq28}
\mu\epsilon\cdot\sqrt{{2r_+}\over{r_+-r_-}}\cdot
\int_{0}^{1}dz\sqrt{{{1}\over{z}}-1}
=\big(k+{1\over2}\big)\cdot\pi\
\ \ \ ; \ \ \ \ k=0,1,2,...\  .
\end{equation}
Interestingly, and most importantly for our analysis, the integral in (\ref{Eq28}) can be evaluated 
analytically to yield the mathematically compact relation \cite{Noteintp}
\begin{equation}\label{Eq29}
\epsilon=
{{\sqrt{2}}\over{\mu}}\cdot\sqrt{{r_+-r_-}\over{r_+}}\cdot\big(k+{1\over2}\big)\
\ \ \ ; \ \ \ \ k=0,1,2,...\  .
\end{equation}
Substituting Eq. (\ref{Eq29}) into Eq. (\ref{Eq24}) and using the relations (\ref{Eq14}) and (\ref{Eq20}) 
one obtains the functional relation \cite{Noteee1,Noteee2}
\begin{equation}\label{Eq30}
\mu_k=\sqrt{{{n}\over{2}}}\cdot\Bigg[1-\sqrt{1-{{Q^2}\over{r^2_+}}}\cdot{{\sqrt{2}}\over{\mu}}
\cdot\big(k+{1\over2})\Bigg]\ \ \ \ ; \ \ \ \ k=0,1,2,...\  .
\end{equation}

It is worth noting that one deduces from Eqs. (\ref{Eq24}) and (\ref{Eq29}) that 
our analysis is valid in the dimensionless large-mass $\mu\gg1$ regime [see Eq. (\ref{Eq15})]. 
Thus, taking cognizance of the analytically derived resonance relation (\ref{Eq30}) one finds that 
our analysis is valid in the large-$n$ regime 
\begin{equation}\label{Eq31}
n\gg1\  .
\end{equation}
Taking cognizance of Eq. (\ref{Eq30}) and the large-$n$ relation (\ref{Eq31}) (which 
implies $\mu=\sqrt{n/2}\cdot[1+O(n^{-1/2})]$) 
one obtains the discrete black-hole-field resonance spectrum 
\begin{equation}\label{Eq32}
\mu_k=\sqrt{{{n}\over{2}}}\cdot\Bigg[1-\sqrt{1-{{Q^2}\over{r^2_+}}}\cdot{{2}\over{\sqrt{n}}}
\cdot\big(k+{1\over2})\Bigg]\ \ \ \ ; \ \ \ \ k=0,1,2,...\  .
\end{equation}

Interestingly, the explicit functional dependence of the dimensionless physical parameter $\mu(n)$ on 
the dimensionless discrete parameter $n$ [or equivalently, on the magnetic charge parameter $P$ 
of the central supporting black hole, see Eq. (\ref{Eq4})] can be deduced 
by substituting into Eq. (\ref{Eq32}) the relations [see Eq. (\ref{Eq4})] 
\begin{equation}\label{Eq33}
Q=\sqrt{{{\kappa}\over{2}}}\cdot{{n}\over{2e}}\ \ \ \ ; \ \ \ \ n\in\mathbb{Z}\
\end{equation}
and [see Eqs. (\ref{Eq14}) and (\ref{Eq32})]
\begin{equation}\label{Eq34}
r_+=\sqrt{{{n}\over{2}}}\cdot{{1}\over{m_{\text{w}}}}\cdot[1+O(n^{-1/2})]\  ,
\end{equation}
in which case one finds the $n$-dependent discrete resonance spectrum 
\begin{equation}\label{Eq35}
\mu_k(n)=\sqrt{{{n}\over{2}}}\cdot\Bigg[1-\sqrt{{{4}\over{n}}-{{\kappa m^2_{\text{w}}}\over{e^2}}}
\cdot\big(k+{1\over2})\Bigg]\ \ \ \ ; \ \ \ \ k=0,1,2,...\
\end{equation}
for the composed 
magnetically-charged-black-hole-linearized-electroweak-field bound-state configurations. 

 
\section{Summary and Discussion}

The recently published physically interesting work \cite{GV} has explicitly proved, using numerical
techniques, that magnetically charged black holes in the composed Einstein-Weinberg-Salam field 
theory (\ref{Eq1}) can support external static matter configurations which are made of spatially regular electroweak fields. 

In particular, it has been revealed in \cite{GV} that the boundary 
between composed magnetic-black-hole-electroweak-field hairy 
configurations and bald magnetic Reissner-Nordstr\"om black holes is 
determined by an $n$-dependent [magnetically-dependent, see Eq. (\ref{Eq4})] 
critical {\it existence-line} $\mu\equiv m_{\text{w}} r_+=m_{\text{w}} r_+(n)$. 
The critical existence-line of the Einstein-Weinberg-Salam theory is composed of 
linearized electroweak field configurations (spatially regular electroweak clouds) which are supported by
central magnetically charged Reissner-Nordstr\"om black holes. 

In the present paper we have used {\it analytical} techniques in
order to explore the physical and mathematical properties of the
composed magnetic-Reissner-Nordstr\"om-black-hole-linearized-electroweak-field cloudy 
configurations in the dimensionless large-mass $\mu\gg1$ regime. 

The main analytical results derived in this paper and their physical implications are as
follows:

(1) We have derived the discrete large-$n$ WKB resonance spectrum [see Eqs. (\ref{Eq5}), 
(\ref{Eq6}), and (\ref{Eq35})] \cite{Noterk}
\begin{equation}\label{Eq36}
\mu_k(n\gg1)=
\sqrt{{{n}\over{2}}}\cdot\Bigg[1-\sqrt{{{4}\over{n}}-{{2{\cal R}^2}\over{\alpha}}}
\cdot\big(k+{1\over2})\Bigg]\ \ \ \ ; \ \ \ \ k=0,1,2,...\
\end{equation}
for the composed magnetically-charged-black-hole-linearized-electroweak-field bound-state configurations, 
where  
\begin{equation}\label{Eq37}
{\cal R}\equiv{{m_{\text{w}}}\over{M_{\text{Pl}}}}\
\end{equation}
is the dimensionless W-boson mass.

(2) Interestingly, the analytically derived resonance formula (\ref{Eq36})
yields the remarkably compact dimensionless expression \cite{Note00}
\begin{equation}\label{Eq38}
[\mu(n\gg1)]_{\text{max}}=
\sqrt{{{n}\over{2}}}\cdot\Bigg(1-\sqrt{{{1}\over{n}}-{{{\cal R}^2}\over{2\alpha}}}\Bigg)\
\end{equation}
for the critical existence-line that characterizes the
Einstein-Weinberg-Salam field theory (\ref{Eq1}) in the large-$n$ regime.

It is worth emphasizing again that the physical significance of the $n$-dependent (magnetically-dependent) 
critical-existence line (\ref{Eq38}) 
in the composed Einstein-Weinberg-Salam field theory stems from the fact that it marks, 
in the large-mass (large-$n$) regime, the boundary
between hairy magnetically-charged-black-hole-electroweak-field bound-state 
configurations and bald magnetic-Reissner-Nordstr\"om 
black-hole spacetimes. 
In particular, for a given value of the discrete magnetic parameter $n$ 
of the central supporting black hole, the hairy
magnetic-black-hole-electroweak-field configurations are
characterized by the critical inequality $\mu(n)\leq [\mu(n)]_{\text{max}}$.

(3) Taking cognizance of the fact that [see Eqs. (\ref{Eq5}) and (\ref{Eq6})] 
\begin{equation}\label{Eq39}
{{{\cal R}^2}\over{2\alpha}}={{\kappa m^2_{\text{w}}}\over{4e^2}}\simeq 1.27\times10^{-33}\ll1\
\end{equation}
one deduces that, in the regime 
\begin{equation}\label{Eq40}
1\ll n\ll {{2\alpha}\over{{\cal R}^2}}\sim10^{33}\  ,
\end{equation}
the analytically derived resonance spectrum (\ref{Eq36}) can be approximated by the remarkably compact functional 
expression 
\begin{equation}\label{Eq41}
\mu_k(1\ll n\ll{{\alpha}/{{\cal R}^2}})=\sqrt{{{n}\over{2}}}\cdot\Bigg[1-{{2}\over{\sqrt{n}}}
\cdot\big(k+{1\over2})\Bigg]\ \ \ \ ; \ \ \ \ k=0,1,2,...\  .
\end{equation}

(4) Finally, it is worth noting that the analytically derived large-$n$ resonant spectrum (\ref{Eq36}) for
the dimensionless mass parameter $m_{\text{w}}r_+$ of the composed 
magnetic-black-hole-linearized-electroweak-field system agrees remarkably well 
with the corresponding exact (numerically computed) resonant 
spectrum of \cite{GV}. 
For example, for $n=100$ one finds from (\ref{Eq36}) the {\it analytically} derived relation 
$[m_{\text{w}}r_+]_{\text{analytical}}=6.364$ which agrees to better than $1\%$ with the exact value 
$[m_{\text{w}}r_+]_{\text{numerical}}=6.426$ \cite{Notemw} as computed {\it numerically} in \cite{GV}. 

\bigskip
\noindent
{\bf ACKNOWLEDGMENTS}
\bigskip

This research is supported by the Carmel Science Foundation. I would
like to thank Yael Oren, Arbel M. Ongo, Ayelet B. Lata, and Alona B.
Tea for helpful discussions.



\begin{thebibliography}{99}

\bibitem{Whee} R. Ruffini and J. A. Wheeler, Phys. Today {\bf 24}, 30
(1971).

\bibitem{Car} B. Carter, in {\it Black Holes}, Proceedings of 1972 Session of Ecole d'ete de Physique Theorique,
edited by C. De Witt and B. S. De Witt (Gordon and Breach, New York, 1973).

\bibitem{Chas} J. E. Chase, Commun. Math. Phys. {\bf 19}, 276 (1970); J. D. Bekenstein,
Phys. Rev. Lett. {\bf 28}, 452 (1972); C. Teitelboim, Lett. Nuovo
Cimento {\bf 3}, 326 (1972); A. E. Mayo and J. D. Bekenstein, Phys.
Rev. D {\bf 54}, 5059 (1996); I. Pena and D. Sudarsky, Class. Quant.
Grav. {\bf 14}, 3131 (1997).

\bibitem{Hart} J. Hartle, Phys. Rev. D {\bf 3}, 2938 (1971); C. Teitelboim, Lett.
Nuovo Cimento {\bf 3}, 397 (1972).

\bibitem{BekVec} J. D. Bekenstein, Phys. Rev. D {\bf 5}, 1239 (1972); {\bf 5}, 2403 (1972);
M. Heusler, J. Math. Phys. {\bf 33}, 3497 (1992); D. Sudarsky,
Class. Quantum Grav. {\bf 12}, 579 (1995).

\bibitem{BekMay} A. E. Mayo and J. D. Bekenstein and, Phys. Rev. D {\bf 54}, 5059 (1996).

\bibitem{Her1n} C. A. R. Herdeiro and E. Radu, Int. J. Mod. Phys. D {\bf 24}, 1542014 (2015).

\bibitem{Hodstationary} S. Hod, Phys. Lett. B {\bf 713}, 505 (2012);
S. Hod, Phys. Lett. B {\bf 718}, 1489 (2013) [arXiv:1304.6474]; S.
Hod, Phys. Rev. D {\bf 91}, 044047 (2015) [arXiv:1504.00009]; 
S. Hod, Phys. Lett. B {\bf 771}, 521 (2017); 
S. Hod, Phys. Rev. D {\bf 96}, 124037 (2017).

\bibitem{fir} M. S. Volkov and D. V. Galtsov, JETP Lett. {\bf 50}, 346 (1989), 
[Pisma Zh. Eksp. Teor. Fiz. {\bf 50} ,312 (1989)]; 
P. Bizo\'n, Phys. Rev. Lett {\bf 64}, 2844 (1990); 
M. S. Volkov and D. V. Galtsov, Sov. J. Nucl. Phys. {\bf 51}, 1171 (1990).

\bibitem{VolGal} M. S. Volkov and D. V. Gal’tsov, Phys. Rept. {\bf 319}, 1 (1999).

\bibitem{hairSky}  S. Droz, M. Heusler, and N. Straumann, Phys. Lett. B {\bf 268}, 371 (1991).

\bibitem{hairHig1} P. Breitenlohner, P. Forgács, and D. Maison, Nucl. Phys. B {\bf 383}, 357 (1992).

\bibitem{hairHig2} B. R. Greene, S. D. Mathur, and C. M. O’Neill, Phys. Rev. D {\bf 47}, 2242 (1993). 

\bibitem{hairstr} P. Kanti, N. E. Mavromatos, J. Rizos, K. Tamvakis, and E. Winstanley, Phys. Rev. D {\bf 54}, 
5049 (1996). 

\bibitem{Hod1} S. Hod, Phys. Rev. D {\bf 86}, 104026 (2012) [arXiv:1211.3202].

\bibitem{Hod2} S. Hod, The Euro. Phys. Journal C {\bf 73}, 2378 (2013)
[arXiv:1311.5298]; S. Hod, Phys. Rev. D {\bf 90}, 024051 (2014)
[arXiv:1406.1179]; S. Hod, Phys. Lett. B {\bf 739}, 196 (2014)
[arXiv:1411.2609]; S. Hod, Class. and Quant. Grav. {\bf 32}, 134002
(2015) [arXiv:1607.00003]; S. Hod, Phys. Lett. B {\bf 751}, 177
(2015); S. Hod, Phys. Lett. B {\bf 758}, 181 (2016) [arXiv:1606.02306]; S. Hod
and O. Hod, Phys. Rev. D {\bf 81}, 061502 Rapid communication (2010)
[arXiv:0910.0734]; S. Hod, Jour. of High Energy Phys. {\bf 01}, 030
(2017) [arXiv:1612.00014]; S. Hod, Phys. Rev. D {\bf 108}, 124028 (2023).

\bibitem{Her1} C. A. R. Herdeiro and E. Radu, Phys. Rev. Lett. {\bf 112}, 221101
(2014); C. L. Benone, L. C. B. Crispino, C. Herdeiro, and E. Radu,
Phys. Rev. D {\bf 90}, 104024 (2014); C. A. R. Herdeiro and E. Radu,
Phys. Rev. D {\bf 89}, 124018 (2014); C. A. R. Herdeiro and E. Radu,
Int. J. Mod. Phys. D {\bf 23}, 1442014 (2014). 

\bibitem{Her2} Y. Brihaye, C. Herdeiro, and E. Radu, Phys. Lett. B {\bf 739}, 1 (2014); J. C.
Degollado and C. A. R. Herdeiro, Phys. Rev. D {\bf 90}, 065019
(2014); C. Herdeiro, E. Radu, and H. R\'unarsson, Phys. Lett. B {\bf
739}, 302 (2014); C. Herdeiro and E. Radu, Class. Quantum Grav. {\bf
32} 144001 (2015); C. A. R. Herdeiro and E. Radu, Int. J. Mod. Phys.
D {\bf 24}, 1542014 (2015); C. A. R. Herdeiro and E. Radu, Int. J.
Mod. Phys. D {\bf 24}, 1544022 (2015); P. V. P. Cunha, C. A. R.
Herdeiro, E. Radu, and H. F. R\'unarsson, Phys. Rev. Lett. {\bf
115}, 211102 (2015); B. Kleihaus, J. Kunz, and S. Yazadjiev, Phys.
Lett. B {\bf 744}, 406 (2015); C. A. R. Herdeiro, E. Radu, and H. F.
R\'unarsson, Phys. Rev. D {\bf 92}, 084059 (2015); C. Herdeiro, J.
Kunz, E. Radu, and B. Subagyo, Phys. Lett. B {\bf 748}, 30 (2015);
C. A. R. Herdeiro, E. Radu, and H. F. R\'unarsson, Class. Quant.
Grav. {\bf 33}, 154001 (2016); C. A. R. Herdeiro, E. Radu, and H. F.
R\'unarsson, Int. J. Mod. Phys. D {\bf 25}, 1641014 (2016); Y.
Brihaye, C. Herdeiro, and E. Radu, Phys. Lett. B {\bf 760}, 279
(2016); Y. Ni, M. Zhou, A. C. Avendano, C. Bambi, C. A. R. Herdeiro,
and E. Radu, JCAP {\bf 1607}, 049 (2016); M. Wang, arXiv:1606.00811.

\bibitem{Gar1} G. Garc\'ia and M. Salgado, Phys. Rev. D {\bf 99}, 044036 (2019); 
G. Garc\'ia and M. Salgado, Phys. Rev. D {\bf 101}, 044040 (2020); 
G. Garc\'ia and M. Salgado, Phys. Rev. D {\bf 104}, 064054 (2021); 
G. Garc\'ia and M. Salgado, arXiv:2307.15888.

\bibitem{Hersc1} C. A. R. Herdeiro, E. Radu, N. Sanchis-Gual, and J. A. Font, Phys. Rev. Lett. {\bf 121}, 101102 (2018).

\bibitem{Hersc2} P. G. S. Fernandes, C. A. R. Herdeiro, A. M. Pombo, E. Radu, and N. Sanchis-Gual,
Class. Quant. Grav. {\bf 36}, 134002 (2019) [arXiv:1902.05079].

\bibitem{Hodsc1} S. Hod, Phys. Lett. B {\bf 798}, 135025 (2019) [arXiv:2002.01948]; 
S. Hod, Phys. Rev. D {\bf 101}, 104025 (2020) [arXiv:2005.10268]; 
S. Hod, The Euro. Phys. Jour. C {\bf 80}, 1150 (2020); 
S. Hod, Jour. of High Energy Phys. {\bf 09}, 140 (2023) [arXiv:2308.12990].

\bibitem{aaa1} D. C. Zou and Y. S. Myung, Phys. Rev. D {\bf 100}, 124055 (2019).

\bibitem{aaa2} P. G. S. Fernandes, Phys. Dark Univ. {\bf 30}, 100716 (2020).

\bibitem{Moh} M. Khodadi, A. Allahyari, S. Vagnozzi, and D. F. Mota, JCAP {\bf 09}, 026 (2020) [arXiv:2005.05992].

\bibitem{GV} R. Gervalle and M. S. Volkov, Phys. Rev. Lett. (2024) [arXiv:2406.14357].

\bibitem{Noteunits} We shall use natural units in which $G=c=1$.

\bibitem{Notenp} We shall assume, without loss of generality, the relation $n\geq0$ which corresponds to $P\geq0$.

\bibitem{Notemap} Note that the dimensionless relation (\ref{Eq11}) maps the semi-infinite 
regime $r\in[r_+,\infty]$ to the infinite regime $y\in[-\infty,\infty]$.

\bibitem{WKB1} L. D. Landau and E. M. Liftshitz, {\it Quantum Mechanics}, 3rd ed.
(Pergamon, New York, 1977), Chap. VII.

\bibitem{WKB2} J. Heading, {\it An Introduction to Phase Integral Methods} (Wiley, New
York, 1962).

\bibitem{WKB3} C. M. Bender and S. A. Orszag, {\it Advanced Mathematical Methods for 
Scientists and Engineers} (McGraw-Hill, New York, 1978), Chap. 10.

\bibitem{Noteintp} Here we have used the integral relation $\int_{0}^{1}dx \sqrt{1/x-1}=\pi/2$.

\bibitem{Noteee1} Here we have used the 
relation $r_+=\gamma/(1+\epsilon)=\gamma\cdot[1-\epsilon+O(\epsilon^2)]$ [see Eq. (\ref{Eq24})]. 

\bibitem{Noteee2} Here we have used the relation $Q^2=r_+r_-$ [see Eq. (\ref{Eq7})].

\bibitem{Noterk} Here we have used the relation $\kappa m^2_{\text{w}}/e^2=2\alpha^{-1}
(m_{\text{w}}/m_{\text{Pl}})^2$ [see Eqs. (\ref{Eq5}) and (\ref{Eq6})]. 

\bibitem{Note00} Note that the critical existence-line of the composed magnetic-black-hole-electroweak-field 
system corresponds to the fundamental ($k=0$) resonant mode of (\ref{Eq36}).

\bibitem{Notemw} Here we have used the numerically computed value $r_+(n=100)=10.29$ of \cite{GV} 
with the relation $m_{\text{w}}=gm_{\text{z}}\simeq\sqrt{0.78}/\sqrt{2}\simeq0.6244$ \cite{GV}, which 
yields the exact (numerically computed) dimensionless value $[m_{\text{w}}r_+]_{\text{num}}(n=100)=6.426$.

\end{thebibliography}
\end{document}